\documentstyle{article}[11pt]\textwidth=160mm\textheight=220mm \oddsidemargin=0pt
\newcommand{\ket}[1]{\mbox{$ | #1 \rangle $}}

\begin{document}

\title{Quantum signature scheme with single photons}

\author{Wang Jian{\footnote{jwang@nudt.edu.cn}}
     \,\,\,\,\,Zhang Quan  \,\,Tang Chao-jing\\
 School of Electronic Science and Engineering,
\\National University of Defense Technology,\\
De Ya Road, Changsha, Hunan 410073, China}
\maketitle

\begin{abstract}
Quantum digital signature combines quantum theory with classical
digital signature. The main goal of this field is to take advantage
of quantum effects to provide unconditionally secure signature. We
present a quantum signature scheme with message recovery without
using entangle effect. The most important property of the proposed
scheme is that it is not necessary for the scheme to use
Greenberger-Horne-Zeilinger states. The present scheme utilizes
single photons to achieve the aim of signature and verification. The
security of the scheme relies on the quantum one-time pad and
quantum key distribution. The efficiency analysis
shows that the proposed scheme is an efficient scheme.\\
{\bf keywords}: {Quantum signature, Quantum one-time pads, Quantum
key distribution}
\end{abstract}


\maketitle
\section{Introduction}
%
%
Quantum cryptography is a cryptographic system using quantum effects
to provide unconditionally secure information exchange. Many
advances have been made in quantum cryptography in recent years,
including quantum key distribution (QKD)\cite{Bennett,B92,E91},
quantum secret sharing\cite{hillery}, quantum authentication and
quantum signature\cite{Zeng,Lee,curty,gottesman,lu1,lu2}. Classical
digital signature is the basis of realizing identity authentication,
data integrity protection and non-repudiation services. Because
classical digital signature is not unconditionally secure, some
quantum signature schemes are proposed in recent
years\cite{Zeng,Lee,gottesman,lu1,lu2}. Quantum digital signature
combines quantum theory with classical digital signature and
utilizes quantum effects to achieve unconditional security.

Gottesman and Chuang proposed a quantum signature scheme based on
quantum one-way functions and quantum Swap-test\cite{gottesman}.
However, their scheme is an inefficient scheme. Zeng and Christoph
proposed an arbitrated quantum signature scheme whose security based
on the correlation of the Greenberger-Horne-Zeilinger (GHZ) triplet
states and the use of quantum one-time pad\cite{Zeng}. Lee et al.
proposed two quantum signature schemes with message
recovery\cite{Lee}. Lee's schemes also use GHZ triplet states, qubit
operations and quantum one-time pad to generate and verify the
signature. L\"{u} and Feng presented two quantum signature schemes,
one scheme based on quantum one-way functions and the other based on
GHZ triplet states and quantum stabilizer codes\cite{lu1,lu2}.

Most of the proposed quantum signature schemes use entangle effect
to achieve the aim of signature and verification. In this paper, we
present a quantum signature scheme with message recovery without
using entangle effect. The communication parties utilize single
photons to perform signature and verification with the help of an
arbitrator. The security of our scheme relies on quantum one-time
pad and quantum key distribution proved as unconditionally
secure\cite{mayers,shor}. Because our scheme does not need to
distribute the GHZ particles and only needs von Neumann measurement,
our scheme provides higher efficiency.
%
\section{Description of our quantum signature scheme}
Our scheme is composed of three parts: initial phase, signature
phase and verification phase, involving three participators, the
signatory Alice, the receiver Bob and the arbitrator. In the initial
phase, the arbitrator distributes secret information to each Alice
and Bob. In the signature phase, Alice generates her signature in
association with the message qubits, the secret information and
qubit operations. In the verification phase, Bob verifies Alice's
signature with the help of the arbitrator.

Both QKD and quantum one-time pad play important roles in our
scheme. We use QKD protocol, such as BB84 or B92 protocol to
distribute the secret information. The encryption algorithm used in
our scheme is quantum one-time pad described in \cite{boykin}. We
select a set of $M$ unitary operations $\{U_k\}$, $k=1,\cdots,M$.
The key $k$ is chosen with probability $p_k$ and the message state
is $\rho$. Then we obtain the cipher state $\rho_c$, where
$\rho_c=\sum\limits_k{p_k{U_k}\rho {U_k^\dag}}=\frac{1}{2^n}I$.

\subsection{Initial Phase}
Alice and Bob share secret keys $K_a$, $K_b$ with the arbitrator,
respectively. These keys are assumed to be distributed via QKD
protocol. The arbitrator generates two random secret bit strings $A$
and $B$, where $A=\{A_1,A_2,\cdots,A_n\}$,
$B=\{B_1,B_2,\cdots,B_n\}$, $A_i,B_i(i=1,2,\cdots,n)\in\{0,1\}$, as
authentication keys and distributes the two strings to each Alice
and Bob via QKD protocol.

\subsection{Signature Phase}
In the signature phase, Alice signs her message $\ket{P}$ and
obtains her signature $\ket{S}$.

Step1: At the beginning, Alice prepares a string of message qubits
$\ket{P}=\{\ket{p_1},\ket{p_2},\cdots,\ket{p_n}\}$, where any qubit
$\ket{p_i}$ ($i=1,2\cdots,n$) in $\ket{P}$ is one of the two
eigenstates $\ket{0}$, $\ket{1}$. $\ket{P}$ may be presented by
classical bits. We represent the message string as qubits so that
the signature can be transmitted through quantum channel.

Step2: In this step, Alice transforms the massage qubits $\ket{P}$
into state $\ket{M}$ according to $A$. If $A_i=0$, Alice does
nothing to $\ket{p_i}$, otherwise she applies bit flip gate
\begin{center}
\[
X=\left(
\begin{array}{cc}
0&1\\
1&0
\end{array}
\right)
\]
\end{center}
to $\ket{p_i}$. Then after Alice's qubit operations, she obtains

\begin{eqnarray*}
\ket{M}=\{\ket{m_1},\ket{m_2},\cdots,\ket{m_n}\}=\{\ket{A_1\oplus
p_1},\ket{A_2\oplus p_2},\cdots,\ket{A_n\oplus p_n}\}.
\end{eqnarray*}

Step3: In this step, Alice transforms the massage qubits $\ket{P}$
into state $\ket{R}$ according to $K_a$. If $K_a^i=0$, Alice uses
rectilinear measurement basis \{\ket{\rightarrow}, \ket{\uparrow}\}.
If $K_a^i=1$, Alice uses diagonal measurement basis
\{\ket{\nearrow}, \ket{\searrow}\}. Then $\ket{0}$, $\ket{1}$ of the
message qubits $\ket{P}$ are represented by \ket{\rightarrow} and
\ket{\uparrow} in the rectilinear measurement basis and by
\ket{\nearrow} and \ket{\searrow} in the diagonal measurement basis,
respectively. Alice obtains
\begin{center}
$\ket{R}=M_{K_a}\ket{P}=\{\ket{r_1},\ket{r_2},\cdots,\ket{r_n}\}$,
\end{center}
where $\ket{r_i}=M_{K_a}^i\ket{p_i}(i=1,2,\cdots,n)$.

Step4: Alice obtains the quantum signature $\ket{S}$ for the message
qubits $\ket{P}$ by encrypting $\ket{M}$ and the secret qubits
$\ket{R}$ with the key $K_a$.
\begin{displaymath}
\ket{S}=E_{K_a}\{\ket{M}, \ket{R}\}.
\end{displaymath}

Step5: Alice sends the signature $\ket{S}$ to Bob through a quantum
channel.

\subsection{Verification phase}
In the verification phase, Bob verifies Alice's signature $\ket{S}$
with the help of the arbitrator.

Step1: After receiving the signature $\ket{S}$, Bob encrypts
$\ket{S}$ and $B$ with the key $K_b$ and obtains
\begin{center}
$\ket{N}=E_{K_b}\{\ket{S},B\}$.
\end{center}
Then he sends $\ket{N}$ to the arbitrator.

Step2: The arbitrator decrypts $\ket{N}$ with the key $K_b$ and
obtains $\ket{S}$, $B'$. Then he decrypts $\ket{S}$ with the key
$K_a$ and obtains $\ket{M}$, $\ket{R}$.

Step3: The arbitrator recovers the message $\ket{P}$ with $\ket{M}$
and $A$ for

\begin{eqnarray*}
\ket{P}=\ket{A\oplus M}=\{\ket{A_1\oplus m_1},\ket{A_2\oplus
m_2},\cdots,\ket{A_n\oplus m_n}\}.
\end{eqnarray*}

The arbitrator then generates $\ket{R_a'}$ using $\ket{P}$ and $K_a$
according to the methods of the third step in the signature phase.
He compares $\ket{R_a'}$ with $\ket{R}$ and generates a verification
parameter $\gamma$. If $\ket{R_a'}=\ket{R}$, he sets $\gamma=0$,
otherwise $\gamma=1$. The arbitrator also compares $B'$ with $B$ and
generates parameter $\xi$. If $B'=B$, he sets $\xi=0$, otherwise
$\xi=1$. The arbitrator transforms the message qubits $\ket{P}$ into
$\ket{R_b'}$, where $\ket{R_b'}=M_{K_b}\ket{P}$, using $K_b$. He
also generates $\ket{U}=\ket{B\oplus P}$ using $B$ and $\ket{P}$.

Step4: The arbitrator encrypts $\gamma$, $\xi$, $\ket{U}$,
$\ket{R_b'}$ and $\ket{S}$ using the key $K_b$ and obtains
\begin{center}
$\ket{V}=E_{K_b}\{\gamma, \xi, \ket{U}, \ket{R_b'}, \ket{S}\}$.
\end{center}
Then he sends $\ket{V}$ to Bob.

Step5: Bob decrypts $\ket{V}$ and obtains $\gamma$, $\xi$,
$\ket{U}$, $\ket{R_b'}$, $\ket{S}$. If $\gamma=\xi=0$, he then
recovers the message qubits by $\ket{P}=\ket{B\oplus U}$. Bob
generates $\ket{R'}$ using $\ket{P}$ and $K_b$, and compares it with
$\ket{R_b'}$. If $\ket{R'}=\ket{R_b'}$, he accepts the message
qubits $\ket{P}$ and the signature $\ket{S}$, otherwise he should
reject it.

%
\section{Security analysis}
The security of signature schemes requires that the signatory can
not disavow her signature and the signature can not be forged. We
demonstrate that our scheme is unconditionally secure as follows.

\subsection{Impossibility of forgery}
\textbf{Theorem 1.} {\it If other entities forge Alice's signature,
their cheating will be detected with a probability $P_r\geq
1-\frac{1}{2^{|K_a|+|A|}}$}.

\textbf{Proof.} The signature is generated by encrypting the state
$\ket{R}$ and $\ket{M}$ with $K_a$, secretly kept by Alice and the
arbitrator. $K_a$ is distributed via QKD protocol proved as
unconditionally secure. $\ket{R}$, $\ket{M}$ are states, into which
Alice transforms the message qubits $\ket{P}$ according to $K_a$ and
$A$, respectively. Firstly, we assume that Bob is dishonest and
tries to forge Alice's signature. However, $\ket{R}$, $\ket{M}$ and
$K_a$ are secret for him. Secondly, we assume an attacker, Eve tries
to forge Alice's signature. The public parameters of our scheme are
$\ket{S}$, $\ket{N}$ and $\ket{V}$, where $\ket{S}=E_{K_a}\{\ket{M},
\ket{R}\}$, $\ket{N}=E_{K_b}\{\ket{S},B\}$,
$\ket{V}=E_{K_B}\{\gamma, \xi, \ket{U}, \ket{R_b'}, \ket{S}\}$ and
they do not offer any information of the secret keys. The encryption
algorithm is quantum one-time pad proved as unconditionally secure.
If Bob or Eve randomly selects the two strings $K_a'$ and $A'$ to
execute the scheme, their cheating will be detected by the
arbitrator with a probability lager than
$1-\frac{1}{2^{|K_a|+|A|}}$, where $|K_a|$ and $|A|$ denote the
length of the $K_a$ and $A$, respectively.

\subsection{Impossibility of disavowal}
\textbf{Theorem 2.}{\it If Alice denies her signature, the
arbitrator can judge whether Alice has disavowed her signature.}

\textbf{Proof.} Because the signature $\ket{S}$ contains Alice's key
$K_a$, Alice can not disavow her signature. If Alice disavows her
signature, Bob only need to send the signature $\ket{S}$ to the
arbitrator and the arbitrator will judge whether Alice has disavowed
her signature. If the signature contains Alice's secret keys, this
signature has been carried by Alice, otherwise, the signature has
been forged by other entities. On the other hand, Bob can not deny
his receiving of Alice's signature because he needs the help of the
arbitrator in the verification phase.

\section{Efficiency analysis}
We simply define the efficiency of quantum signature schemes as the
formula $\eta=B_s/(Q_t+B_t)$, where $B_s$ is the number of message
qubits signed, $Q_t$ and $B_t$ are each the number of qubit
transmitted and classical bit exchanged in the scheme. In terms of
the formula, the efficiency of our scheme, Lee's scheme and Zeng's
scheme is each 11\%, 11\% and 9\%. However, the formula is just used
to simply analyze the efficiency of quantum signature schemes. In
practice, we should also consider the complexity of realizing a
scheme, involving the distribution of initial secret information,
measurement methods, encryption algorithm, etc.

Although an arbitrator is not necessary in Gottesman's scheme, it
requires a trusted key distribution center, which has authenticated
links to all three participators. Moreover, $M$ keys are used to
sign each bit message in Gottesman's scheme. Zeng's scheme requires
a joint measurement on each message qubit and GHZ particle to
generate a four-particle entangled state and $n$ Bell measurement to
disentangle message qubits and GHZ particles. In Lee's scheme, the
signatory and the verifier should also measure their GHZ particles
and compare qubits strings in the verification phase. L\"{u}'s
scheme is relatively complicated, because the secret keys and the
syndromes of quantum stabilizer codes are each used to encrypt and
encode the quantum states. Our scheme does not need to use the GHZ
triplet states, so it reduces the process of distributing and
measuring the GHZ particles. In the verification phase, we only need
to use von Neumann measurement to verify the qubits strings.
Compared with these quantum signature schemes, our scheme provides
higher efficiency.

\section{Conclusion}
In this paper, we propose a quantum signature scheme with single
photons. In the initial phase of our scheme, secret keys and
authentication keys are distributed to each participator via QKD
protocol. In the signature phase, Alice generates her signature by
encrypting the transformed message qubits with the secret keys. In
the verification phase, Bob verifies Alice's signature with the help
of the arbitrator. The security analysis showed that our scheme is
unconditionally secure. We only need to utilize QKD, quantum
one-time pad, qubit operations and von Neumann measurement to
realize our scheme, so the scheme is simple and practicable. Many
potential improvements of quantum signature remain largely
unexplored. The properties of quantum should be fully used to design
more practicable quantum signature schemes.


\section*{Acknowledgments}
This work is supported by the National Natural Science Foundation of
China under Grant No. 60472032.

%
%

%
%
\end{document}